\documentclass[aps,prc,twocolumn,superscriptaddress,showpacs,floatfix]{revtex4-1}

\usepackage{graphicx,amssymb}
\usepackage[fleqn]{amsmath}
\usepackage{amsfonts}
\usepackage{dcolumn}
\usepackage{bm}
\usepackage{braket}
\usepackage{multirow} 
\usepackage{MnSymbol}
\usepackage{hyperref}
\hypersetup{letterpaper, colorlinks=true, urlcolor=blue, citecolor=blue, linkcolor=blue, breaklinks=true}
\usepackage{nccmath}

\graphicspath{{images/}}

\begin{document}

\date{\today}

\title{Search for excited states in $^{25}$O}

\author{M.D.~Jones}
\affiliation{Nuclear Science Division, Lawrence Berkeley National Laboratory, Berkeley, CA 94720, USA}
\author{K.~Fossez}
\affiliation{National Superconducting Cyclotron Laboratory, Michigan State University, East Lansing, MI 48824, USA}
\affiliation{Facility for Rare Isotope Beams, Michigan State University, East Lansing, MI 48824, USA}
    \author{T.~Baumann}
    \affiliation{National Superconducting Cyclotron Laboratory, Michigan State University, East Lansing, MI 48824, USA}
   \author{P.A.~DeYoung}
   \affiliation{Department of Physics, Hope College, Holland, MI 49422-9000, USA}
    \author{J.E.~Finck}
    \affiliation{Department of Physics, Central Michigan University, Mount Pleasant, MI 48859, USA}
    \author{N.~Frank}
    \affiliation{Department of Physics and Astronomy, Augustana College, Rock Island, IL 61201, USA}
    \author{A.N.~Kuchera}
    \affiliation{Department of Physics, Davidson College, Davidson, NC 28035, USA}
    \author{N.~Michel}
    \affiliation{Facility for Rare Isotope Beams, Michigan State University, East Lansing, MI 48824, USA}
   \author{W. Nazarewicz}
   \affiliation{Facility for Rare Isotope Beams, Michigan State University, East Lansing, MI 48824, USA}
   \affiliation{Department of Physics and Astronomy, Michigan State University, East Lansing, MI 48824, USA} 
    \author{J.~Rotureau}
    \affiliation{Facility for Rare Isotope Beams, Michigan State University, East Lansing, MI 48824, USA}
    \author{J.K.~Smith}
    \affiliation{Department of Physics, Reed College, Portland, OR 97202-8199, USA}
    \author{S.L.~Stephenson}
    \affiliation{Department of Physics, Gettysburg College, Gettysburg, PA 17325, USA}
    \author{K.~Stiefel}
    \affiliation{National Superconducting Cyclotron Laboratory, Michigan State University, East Lansing, MI 48824, USA}
    \affiliation{Department of Chemistry, Michigan State University, East Lansing, MI 48824, USA}
       \author{M.~Thoennessen}  
   \affiliation{National Superconducting Cyclotron Laboratory, Michigan State University, East Lansing, MI 48824, USA}
   \affiliation{Department of Physics and Astronomy, Michigan State University, East Lansing, MI 48824, USA}
    \author{R.G.T.~Zegers}
    \affiliation{National Superconducting Cyclotron Laboratory, Michigan State University, East Lansing, MI 48824, USA}
    \affiliation{Department of Physics and Astronomy, Michigan State University, East Lansing, MI 48824, USA}
    \affiliation{Joint Institute for Nuclear Astrophysics -- Center for the Evolution of the Elements, Michigan State University, East Lansing, Michigan 48824, USA}

\begin{abstract}
  \begin{description}
    \item[Background]
    Theoretical calculations suggest the presence of low-lying excited states in $^{25}$O. Previous experimental searches by means of proton knockout on $^{26}$F produced no evidence for such excitations.
    \item[Purpose]
      We search for excited states in $^{25}$O using the ${ {}^{24}\text{O} (d,p) {}^{25}\text{O} }$ reaction. The theoretical analysis of excited states in unbound $^{25,27}$O is based on the  configuration interaction approach that accounts for  couplings to the scattering continuum.
    \item[Method]
      We use invariant-mass spectroscopy to measure neutron-unbound states in $^{25}$O. For the theoretical approach, we use the complex-energy Gamow Shell Model and Density Matrix Renormalization Group method with a finite-range two-body interaction optimized to the bound states and resonances of $^{23-26}$O, assuming a core of $^{22}$O. We predict energies, decay widths, and asymptotic normalization coefficients.
    \item[Results]
      Our calculations in a large $spdf$ space predict several low-lying excited states in $^{25}$O of positive and negative parity, and we obtain an experimental limit on the relative cross section of a possible ${ {J}^{\pi} = {1/2}^{+} }$ state with respect to the ground-state of $^{25}$O at $\sigma_{1/2+}/\sigma_{g.s.} = 0.25_{-0.25}^{+1.0}$. We also discuss how the observation of negative parity states in $^{25}$O could guide the search for the low-lying negative parity states in $^{27}$O.
    \item[Conclusion]
      Previous experiments based on the proton knockout of $^{26}$F suffered from the low cross sections for the population of excited states in $^{25}$O because of low spectroscopic factors. In this respect, neutron transfer reactions carry more promise. 
  \end{description}
\end{abstract}

\maketitle

\section{Introduction}
{
  In recent years, the search for excited states in $^{25}$O has led to conflicting results between experimental and theoretical works, calling for a re-examination. Recent experimental results \cite{hoffman08_1781,caesar13_1765,kondo16_1439} from proton knockout on $^{26}$F established the unbound ground state of $^{25}$O as a ${ {J}^{\pi} = {3/2}^{+} }$ neutron resonance at  $ \approx 0.75$\,MeV relative to $^{24}$O, with a width of ${ \Gamma \approx 90 }$ keV. However, none of these studies could identify any excited states.

  In contrast, various theoretical approaches predict the presence of excited states in $^{25}$O. 
  For instance, the USDB shell model (SM) \cite{brown17_1876} predicts an excited state with ${ {J}^{\pi} = {1/2}^{+} }$ at about 3.3 MeV above the ground state (g.s.), 
  in relative agreement with the continuum SM (CSM) \cite{volya14_1478} and the Gamow shell model (GSM) \cite{sun17_1840,fossez17_1927}. 
  The CSM and the GSM also predict another narrow state with ${ {J}^{\pi} = {5/2}^{+} }$ at  ${ E \approx 4.5 }$ MeV. 
  These states are expected to have dominant configurations with a hole in the neutron shell $1s_{1/2}$, which would make them difficult to populate in proton knockout reactions. 

  For the negative-parity states, the situation is also unclear because of the inconsistent predictions between various theoretical approaches \cite{hagino16_1848,hagen16_1800}. 
  Negative parity states might be present in the low-energy spectrum of $^{25}$O because of the increasing couplings to the $fp$ continua as one approaches the neutron drip-line. 
  This can be illustrated by plotting the Hartree-Fock-Bogoliubov canonical energies 
  as a function of the neutron number for $^{18-28}$O as shown in Fig.~\ref{fig1}.
  The chemical potential is also shown to indicate which shells are occupied for a given isotope. 
  %
  \begin{figure}[htb]
    \includegraphics[trim=5 12 0 0, clip, width=1.0\linewidth]{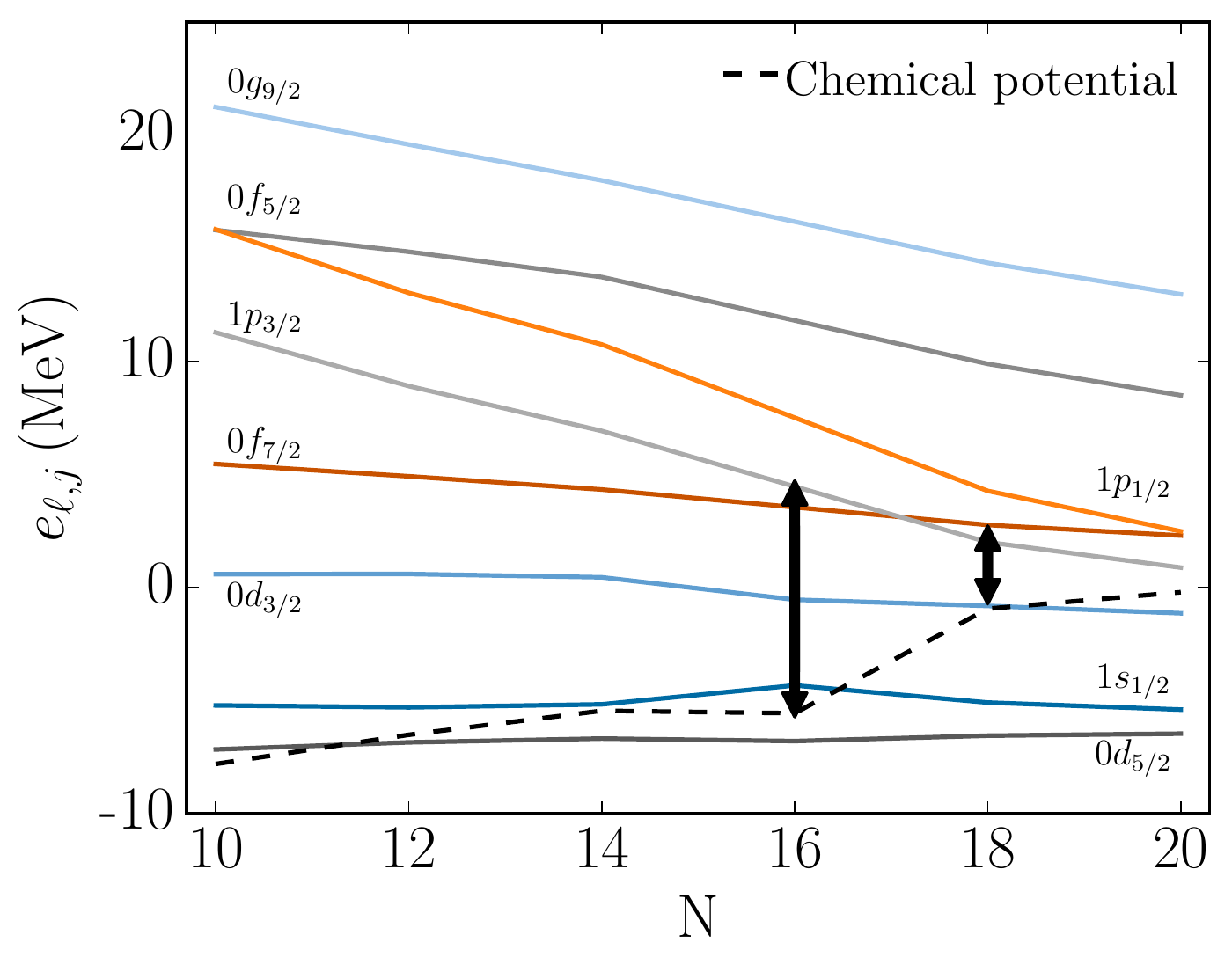}
    \caption{Hartree-Fock-Bogoliubov (HFB) canonical energies as a function of the neutron number for oxygen isotopes. The calculations with
     the energy density functional UNEDF0 \cite{kortelainen10_1928} were performed by means of the HFB solver \cite{stoitsov08_139} based on
    P{\"o}schl-Teller-Ginocchio and Bessel/Coulomb wave functions,
    which properly takes the continuum couplings into account. The chemical potential (in MeV) is represented by a dashed line. While the ground state of $^{28}$O is bound in these calculations, it does not significantly affect the picture. The arrows show the decreasing energy gap between the last occupied level and the $fp$ levels in $^{24}$O and $^{26}$O, respectively.}
    \label{fig1}
  \end{figure}
  While the $sd$-shell energies stay almost constant as the mass number increases, 
  the canonical energies of the $fp$ shells decrease because of the increasing couplings to the $fp$ neutron continuum. 
  This effect is even more pronounced when the gap between the chemical potential and the last occupied shell decreases sharply between $^{24}$O and $^{26}$O, 
  because of the progressive filling of the ${ 0{d}_{3/2} }$ shell as shown in Fig.~\ref{fig1}. 

  A dramatic consequence of this reduced gap, resulting in a clustering of canonical states around the neutron emission threshold,  are the strong dineutron correlations in the g.s.~of the two-neutron emitter $^{26}$O \cite{guillemaud90_1761,lunderberg12_556,kohley13_1541,caesar13_1765,kondo16_1439}, 
  which cannot be explained without the admixture of positive and negative parity states in the wave function \cite{fossez17_1927,catara84_1880,pillet07_1879,hagino14_1160,hagino16_1881}.
  The phenomenon described above is not limited to the neutron-rich oxygen isotopes, 
  and simply reflects the transition from a mean-field-dominated to correlation-dominated regime as one approaches the drip-line and couplings to the continuum increase \cite{dobaczewski96_1950,fayans00_1948,meng02_1949,dobaczewski07_17,zhang11_1951,forssen13_394,dobaczewski13_b244}.
While several theoretical studies investigated the role of the continuum in neutron-rich oxygen isotopes \cite{tsukiyama09_497,hagen12_685,tsukiyama15_1442}, 
  only few provided insight into the couplings to negative-parity continuum states \cite{hagino16_1848,hagen16_1800,fossez17_1927}. 

  In this work, we present the first experimental results for $^{25}$O obtained in a neutron-transfer experiment, in which states previously inaccessible in proton knockout can be observed. In addition, we re-examine the existence of excited states in $^{25}$O by considering couplings to the $fp$ continuum and make a case for future experimental studies.
  In Sec.~\ref{sec_exp} we present the experimental results obtained for the ${ {}^{24}\text{O} (d,p) {}^{25}\text{O} }$ transfer reaction 
  and provide a limit on the relative cross section for a possible positive-parity excited state in $^{25}$O. 
  The theoretical predictions for both positive- and negative- parity states 
  in $^{25}$O are presented in Sec.~\ref{sec_th}, with a discussion of the consequences for excited states in $^{27}$O. 

}

\section{Experimental method}
{
  \label{sec_exp}

  The experiment was conducted at the National Superconducting Cyclotron Laboratory (NSCL), where a secondary beam of $^{24}$O was produced by fragmentation of $^{48}$Ca on a $^{9}$Be target. The A1900 fragment separator was used to separate $^{24}$O from the other fragmentation products and transport it to the experimental area at an energy of 83.4 MeV/nucleon where the $^{24}$O impinged upon the Ursinus College Liquid Hydrogen Target, filled with liquid deuterium. The remaining beam contaminants were removed by time-of-flight (ToF) in the off-line analysis. 
 
  A $(d,p)$ transfer reaction on $^{24}$O was used to populate neutron-unbound states in $^{25}$O, which promptly decayed. The reacted $^{24}$O was swept 43.3$^{\circ}$ by a 4-Tm superconducting sweeper magnet \cite{sweeper} into a collection of position- and energy-sensitive charged-particle detectors, where the position and momentum at the target were reconstructed with an inverse transformation matrix \cite{nathan_nim,cosy}. Element separation was accomplished through energy-loss and ToF, and isotope identification was obtained through correlations in the ToF, dispersive position, and dispersive angle. This technique is described in further detail in Ref. \cite{GregPRC2012}.

  The neutrons emitted in the decay of $^{25}$O traveled forwards toward the Modular Neutron Array (MoNA) \cite{mona_nimA} and the Large-area multi-Institutional Scintillator Array (LISA). The momentum vectors of the incident neutrons were determined from their location in MoNA-LISA and ToF. Together, the sweeper and MoNA-LISA provide a complete kinematic measurement of the recoiling $^{24}$O and the neutron, allowing the decay to be reconstructed. Additional information on the experimental setup can be found in Ref. \cite{jones15_1351,MDJThesis}. 
  The two-body decay energy of $^{25}$O  is 
 $E_{\mathrm{decay}} = M^{*} - M_{\mathrm{^{24}O}} - m_{n}$,
where $M^{*}$ is the invariant mass of the decaying system, $M_{\mathrm{^{24}O}}$ the mass of $^{24}$O and $m_{n}$ the mass of the neutron. The invariant mass of the two-body system is obtained from the experimentally measured four-momenta of $^{24}$O and the first time-ordered interaction in MoNA-LISA. 

 \begin{figure}[ht!]
    \includegraphics[trim=0 0 0 20, clip, width=1.0\linewidth]{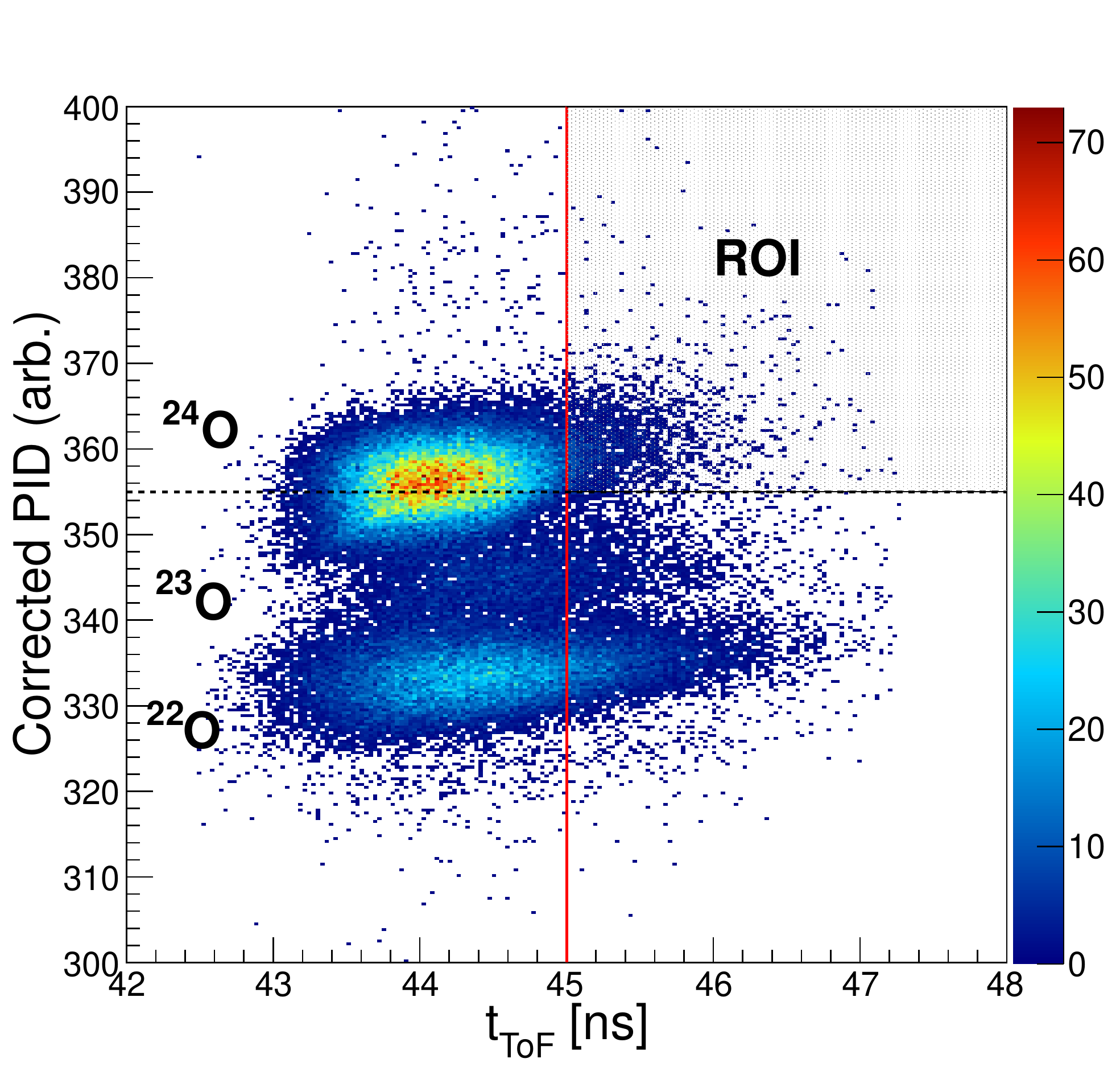}
    \caption{Corrected Particle Identification (PID) vs. time-of-flight in the sweeper ($t_{\text{ToF}}$) for the oxygen isotopes. The black-dashed line denotes the selection of $^{24}$O, and the red-solid line the selection on time-of-flight. The $(d,p)$ reaction products are expected to be populated in the region of interest (denoted ROI, shaded grey).}
    \label{fig2a}
  \end{figure}

  In this experiment the neutron transfer to unbound states in $^{25}$O introduces an additional challenge as the fragments of interest are identical to the unreacted beam apart from their energy. In this respect, isotope separation is insufficient. In order to isolate the reaction products and reduce the background from the unreacted beam, a gate on late ToF from the target to the end of the sweeper of $ t_{\text{ToF}} > 45 $ ns was applied to the $^{24}$O fragments. This selection is shown in Fig. \ref{fig2a}, where the Particle Identification (PID) is shown vs. the time-of-flight. One expects the reaction products to deviate from the beam spot due to the dynamics of the $(d,p)$ reaction and the subsequent neutron evaporation. A similar selection was made in a previous $(d,p)$ experiment with MoNA-LISA to isolate the reaction products \cite{JessePRC}.
To further isolate the reaction products, a coincidence in MoNA-LISA was required with a threshold of 2 MeV of equivalent electron energy (MeVee) and a ToF gate on prompt neutrons was also applied.

  While these requirements significantly reduce the background from unreacted beam, they do not eliminate it. For this reason, a background measurement was taken on an empty target. Figure \ref{fig2} shows the measured neutron kinetic energy for the full- (blue-solid) and empty- (grey-dashed) targets with identical cuts except for the fragment ToF (due to the change in rigidity). The empty-target data are scaled to the integral of the full-target data for kinetic energies below KE$_{n} < 45$ MeV/nucleon, where the spectrum is dominated by background. In the inset, the background-subtracted kinetic energy is shown where a peak at the center-of-target energy is evident. Only events above KE$_{n} > 45$ MeV/nucleon are included in the two-body decay energy, where the remaining background contamination is $30\%$, with $98\%$ of the neutron kinetic energy distribution included.

  \begin{figure}[ht!]
    \includegraphics[trim=5 0 30 20, clip, width=1.0\linewidth]{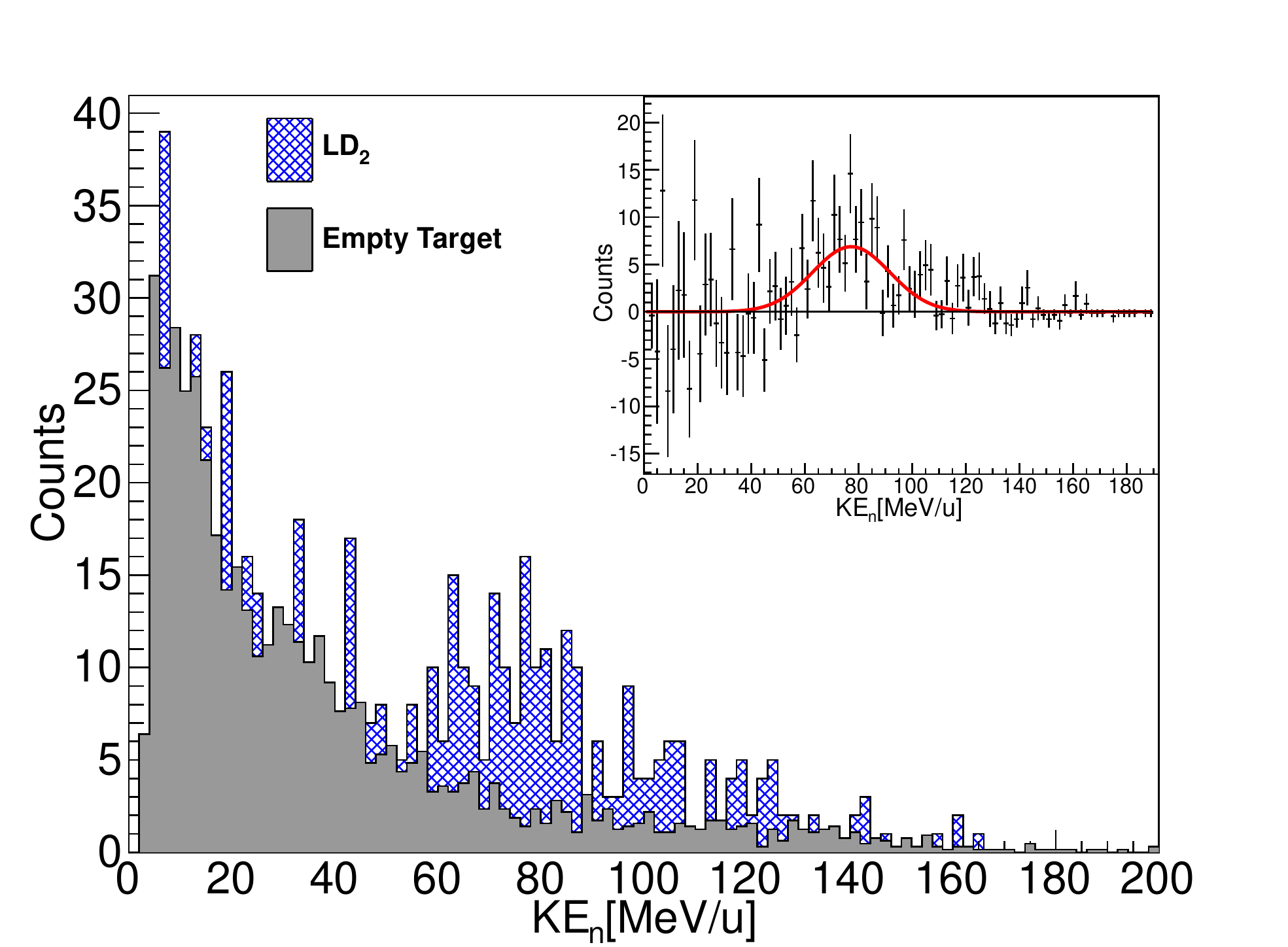}
    \caption{ Neutron kinetic energy for the full liquid deuterium target (hatched-blue) gated in the ROI of Fig. \ref{fig2a}, and empty target measurement (solid-grey). The inset shows the background subtracted spectrum, with a Gaussian fit (solid-red).}
    \label{fig2}
  \end{figure}

  The two-body decay energy for $^{25}$O can be found in Fig.~\ref{fig3}, where a peak at the previously reported ground-state energy from proton-knockout \cite{hoffman08_1781,caesar13_1765,kondo16_1439} is observed in addition to a broad tail. To model the decay, a Monte Carlo simulation incorporating the transfer kinematics, beam characteristics, and subsequent decay and transport through the experimental apparatus was used. The efficiency and acceptance of the charged-particle detectors along with the response of MoNA are fully incorporated into the simulation making the result directly comparable to experiment. The neutron interactions in MoNA were modeled with GEANT4 \cite{GEANT} and MENATE\_R \cite{MENATER}. 

For resonant contributions, the input decay energy line shape was assumed to be of the Breit-Wigner form:
  \begin{ceqn}
    \begin{equation}
      \sigma_{\ell}(E) \sim \frac{\Gamma_{l}}{(E_{0} - E + \Delta_{l})^{2} + \frac{1}{4}\Gamma_{\ell}^{2}},
      \label{eq_crosssec}
    \end{equation}
  \end{ceqn}
where $E_{0}$ is the peak position, $\Delta_{\ell}$ the resonance shift, $\Gamma_{\ell}$ the energy-dependent width.
The largest background contribution is from accidental coincidences with the unreacted beam. The two-body decay energy line-shape from the unreacted beam was determined using the empty-target data and fixed at $30\%$ the total integrated counts to equal its contribution in the kinetic energy (Fig. \ref{fig2}). This component is shown in green (dash-dot-dot) in Fig.~\ref{fig3}.

  \begin{figure}[ht!]
    \includegraphics[trim=5 0 20 15, clip, width=1.0\linewidth]{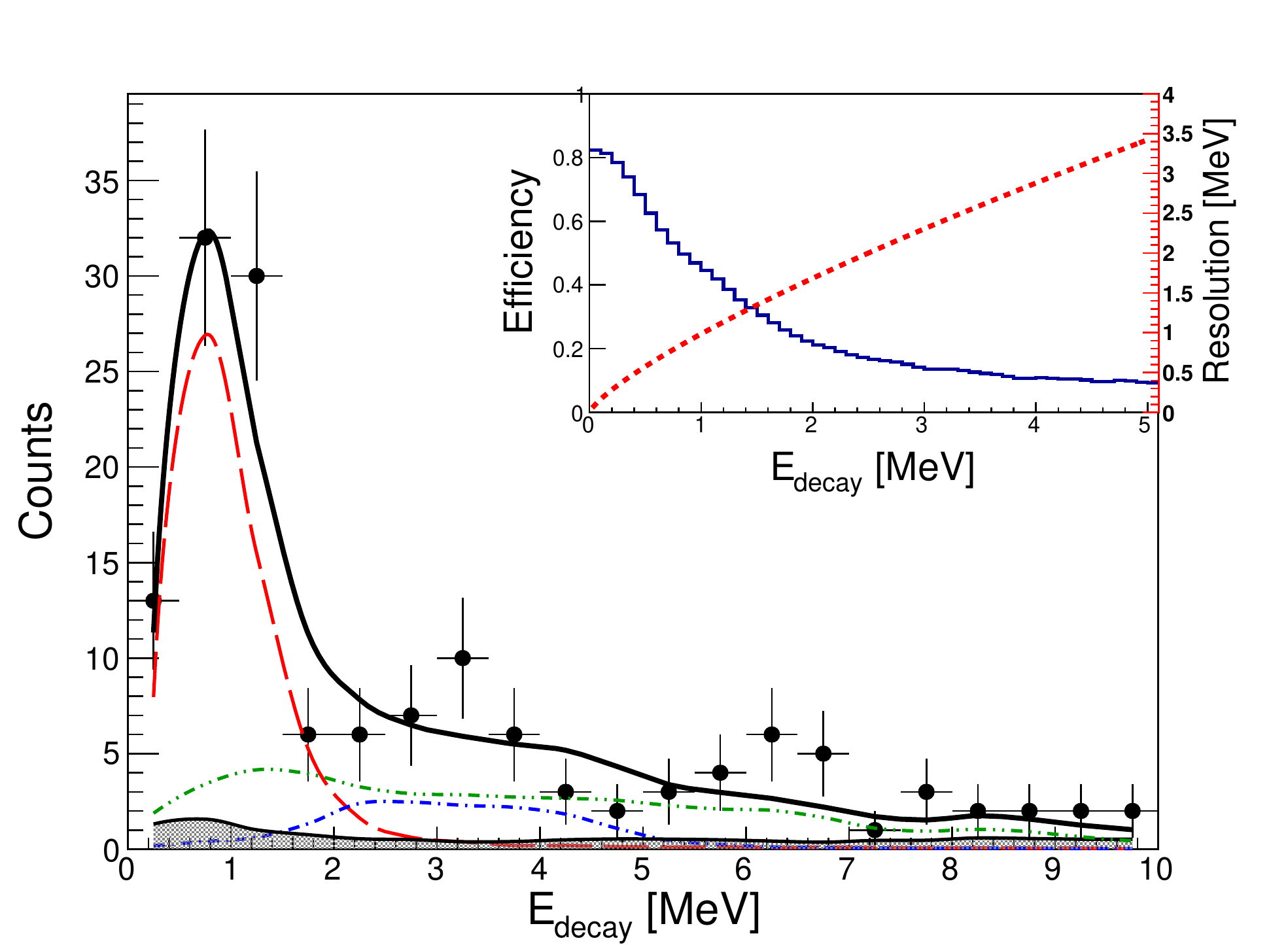}
    \caption{ Two-body decay energy for $^{24}$O + $1n$. The best fit includes a 830 keV resonance (dashed-red), a 3.3 MeV resonance from a possible $J=1/2^{+}$ state at $\sigma_{1/2+}/\sigma_{g.s.} = 1.0$ (dot-dash, blue), and background contributions from the unreacted beam (dash-dot-dot green), in addition to deuteron breakup (shaded-grey). The sum of all components is in solid-black. 
  }
  \label{fig3}
\end{figure}

Contributions from deuteron breakup were also considered. Since the excited states of $^{24}$O are unbound, this was modelled as inelastic excitation of the deuteron, $^{24}$O$(d,d^{*})^{24}$O.
The resulting neutron from the dissociation of the deuteron was boosted into the lab frame and paired artificially with an $^{24}$O event in the simulation to construct a two-body decay energy. Given the acceptance of MoNA-LISA, neutrons coming from deuteron breakup are expected to arrive with velocities between $\beta \sim 0.5 - 0.65$ and $\beta < 0.05$. Since this significantly deviates from the center-of-target beam velocity ($\beta_{beam} \sim 0.38$), the decay energy spectrum peaks at excessively large energies. Nevertheless, the spectrum has a low-energy tail that may still contribute to $E_{\rm decay} < 10$ MeV. It is therefore included as an additional component in the fitting procedure.

The present data are sufficiently described by a single $\ell$=2 resonance at $E = 830 \pm 170$ keV and are insensitive to the width. The best-fit energy agrees well with previous measurements: $770^{+20}_{-10}$ keV \cite{hoffman08_1781}, $725^{+54}_{-29}$ keV \cite{caesar13_1765}, and $749(10)$ keV \cite{kondo16_1439}. There does not appear to be any additional strength below $E_{\rm decay} = 1$ MeV, confirming the assignment of this state as the ground-state. Due to the increased efficiency at low decay energies, it is unlikely for another resonance to be present in this region. 

There is a small excess of counts around $E_{\rm decay} \sim 3.5$ MeV 
where recent GSM calculations \cite{fossez17_1927} predict narrow $1/2^{+}$, $5/2^{+}$ and ${7/2}^{-}$ resonances. 
The dominant configuration of the positive parity states is built on a hole in $1s_{1/2}$ neutron shell, while the negative parity state has a large contribution from the $fp$ continuum.

For these reasons, these states have a small overlap with the g.s. of $^{26}$F and hence they would not be populated in the knockout reaction. 
In principle, the $(d,p)$ reaction can populate these states, 
however the present data are insufficient to confirm the observation of any state in this region, and the $J^{\pi}$ is tentative based on the theoretical interpretation. Only a limit can be determined.
The data are consistent with the inclusion of a first-excited $1/2^{+}$ state at $E = 3.3$ MeV and $\Gamma= 1$ keV, 
with a relative cross-section of $\sigma_{1/2+}/\sigma_{g.s.} = 0.25_{-0.25}^{+1.0}$ with respect to the ground-state. The error on the cross-section ratio is purely statistical and determined by $\chi^{2}$ analysis. While higher-lying states may be present, the experimental data are dominated by background contributions above $E_{\rm decay} = 5$ MeV, and they cannot be resolved. 

}

\section{Theoretical analysis}
{
  \label{sec_th}

  \begin{figure}[ht!]
    \includegraphics[width=1.0\linewidth]{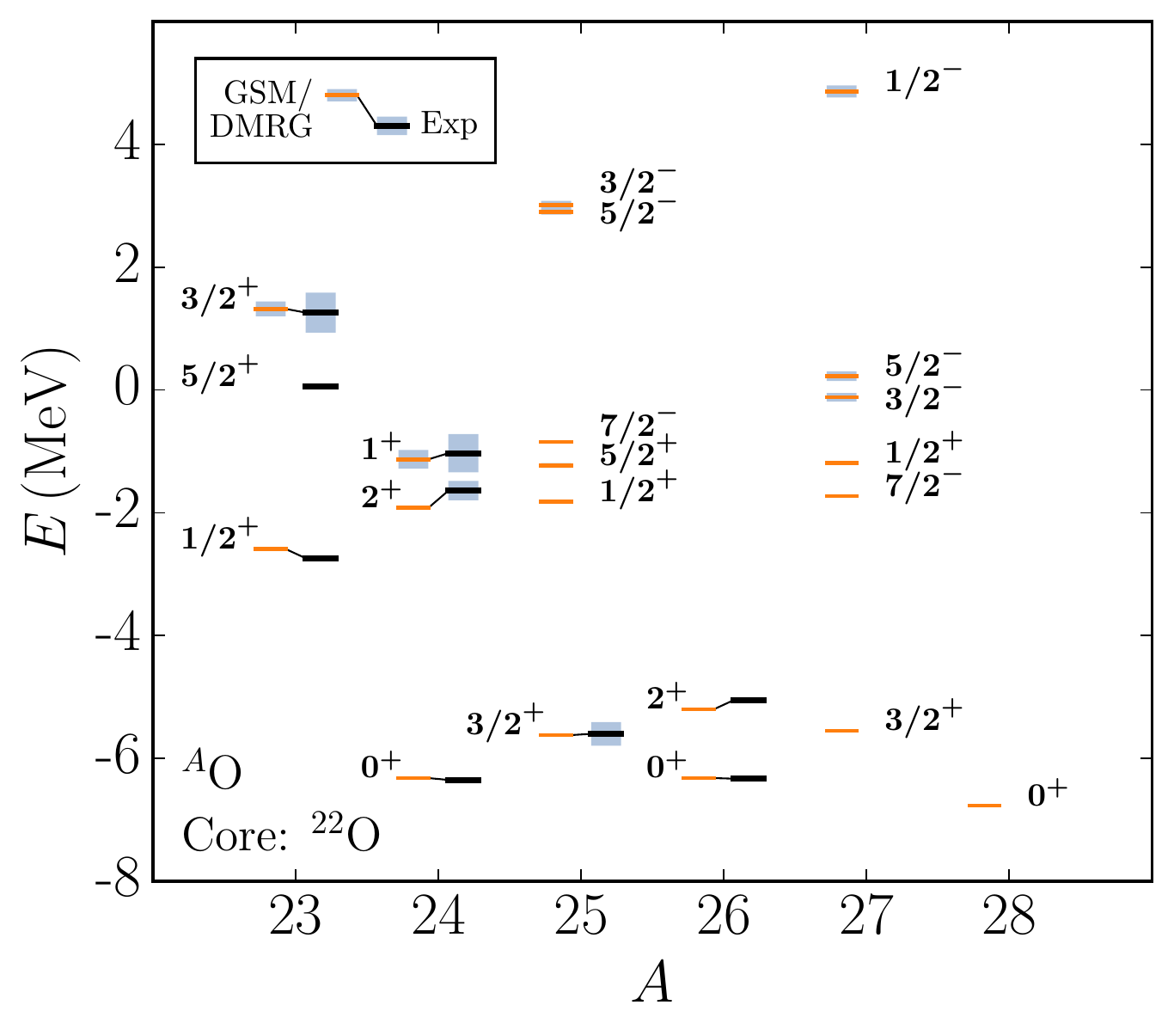}
    \caption{Energies of $^{23-28}$O obtained in the GSM and DMRG approaches compared to experiment. The widths are marked by shaded bands.}
    \label{fig4}
  \end{figure}

  In the present work, we investigate the excited states of $^{25}$O in the GSM framework \cite{michel09_2} by using a core of $^{22}$O and optimizing an effective two-body interaction to the experimentally known states in ${ {}^{23-26}\text{O} }$. 
The GSM is a complex-momentum generalization of the traditional shell model 
  through the use of the Berggren basis \cite{berggren68_32,berggren82_28}. 
    The Berggren basis is defined for each partial wave ${ c = (\ell, j) }$ for which the continuum is expected to be important in the problem at hand, 
  and is made of single-particle bound states, decaying resonances and nonresonant scattering states. 
    The many-body Schr\"odinger equation can be solved either by diagonalizing the Hamiltonian matrix  
  or, if this is is not feasible,  by using the density matrix renormalization group (DMRG) method \cite{white92_488,rotureau06_15,rotureau09_140}. 
  In the DMRG method, continuum couplings are included progressively to an initial wave function obtained in a truncated many-body space by adding scattering states one-by-one, and retaining the many-body states that have large contributions to the GSM density matrix.

  The details of our implementation of the GSM/DMRG framework strictly follow Ref.~\cite{fossez17_1927}.  In particular, the parameters of the core-valence  potential representing the interaction between  $^{22}$O  and the valence neutrons was optimized to   the single-particle states of $^{23}$O. 
The effective two-body interaction used is the finite-range Furutani-Horiuchi-Tamagaki force \cite{furutani78_1012,furutani79_1013}, 
  which has central, spin-orbit, tensor and Coulomb terms. 
  The optimization of the interaction was performed in Ref.~\cite{fossez17_1927} 
  where we demonstrated that the ${sd}$ space was not sufficient to describe the g.s. of $^{26-28}$O, and the ${fp}$-continuum  was essential
to provide the necessary couplings for the description of $^{26-28}$O.
  Hereafter we use the two-body interaction reoptimized for the $spdf$ space to describe excited states in $^{25}$O.
 Our predictions for negative parity states are shown in Fig.~\ref{fig4} together with results from Ref.~\cite{fossez17_1927}.

  For $^{25}$O we predict a ${ {J}^{\pi} = {7/2}^{-} }$ state close to the excited positive parity states ${ {J}^{\pi} = {1/2}^{+} }$ and ${ {5/2}^{+} }$ 
  already predicted in other approaches, as well as two negative parity states ${ {J}^{\pi} = {5/2}^{-} }$ and ${ {3/2}^{-} }$ at higher energy. 
  Interestingly the same pattern is predicted for $^{27}$O, 
  but the negative parity states move down in energy, 
  and a ${ {J}^{\pi} = {1/2}^{-} }$ state is predicted at 10.4\,MeV as shown in Table~\ref{tab_E}. 
  This echoes the pattern described in Fig.~\ref{fig1} wherein the negative parity states move down in energy with the neutron number. 
  As a consequence, the possible observation of negative parity states in $^{25}$O would offer us hints about  such states  in $^{27}$O.
  \begin{table}[htb]
    \caption{Predicted energies (in MeV) and widths (in keV; in parenthesis) in $^{25,27}$O. The separation energy of the g.s. of $^{25,27}$O with the g.s. of $^{24}$O is indicated at the bottom. It is important to note that while the energies of excited states in $^{25}$O are fairly robust with respect to changes of the GSM interaction parameters, those in $^{27}$O exhibit appreciable variations.}
    \begin{ruledtabular}
      \begin{tabular}{cccc}
	${ {J}^{\pi} }$		& ${ E ({}^{25}\text{O}) }$	& ${ {J}^{\pi} }$	& ${ E ({}^{27}\text{O}) }$ \\
	\hline \\[-6pt]
	${ {3/2}^{+} }$	& 0.0 (51)			& ${ {3/2}^{+} }$	& 0.0 (0) \\
	${ {1/2}^{+} }$	& 3.80 (0)			& ${ {7/2}^{-} }$	& 3.82 (0) \\
	${ {5/2}^{+} }$	& 4.39 (79)			& ${ {1/2}^{+} }$	& 4.36 (0) \\
	${ {7/2}^{-} }$	& 4.77 (15)			& ${ {3/2}^{-} }$	& 5.43 (139)\\
	${ {5/2}^{-} }$	& 8.52 (89)			& ${ {5/2}^{-} }$	& 5.78 (157) \\
	${ {3/2}^{-} }$	& 8.63 (137)			& ${ {1/2}^{-} }$	& 10.4 (194) \\
	\hline \\[-6pt]
	${ {E}_{\text{g.s.}} - {E}_{\text{g.s.}} ({}^{24}\text{O}) }$	& 0.70	& 	& 0.77
      \end{tabular}
    \end{ruledtabular}
    \label{tab_E}
  \end{table}

  In order to inform future experimental studies, 
  we calculated the asymptotic normalization coefficients (ANCs) \cite{okolowicz12_475} 
  between states in $^{24}$O and $^{25}$O. 
As shown in Fig.~\ref{fig5}, only the ANC between the ground states of $^{24}$O and $^{25}$O and the ANCs between the excited positive parity states of $^{24}$O and those in $^{25}$O are significant. 
  The imaginary part of the ANC can be interpreted as the uncertainty on the real part for unbound states. In the future, the extension of the GSM formalism to the description of $(d,p)$ reactions could help in the identification of possible excited states in $^{25}$O.
  \begin{figure}[htb]
    \includegraphics[width=1.0\linewidth]{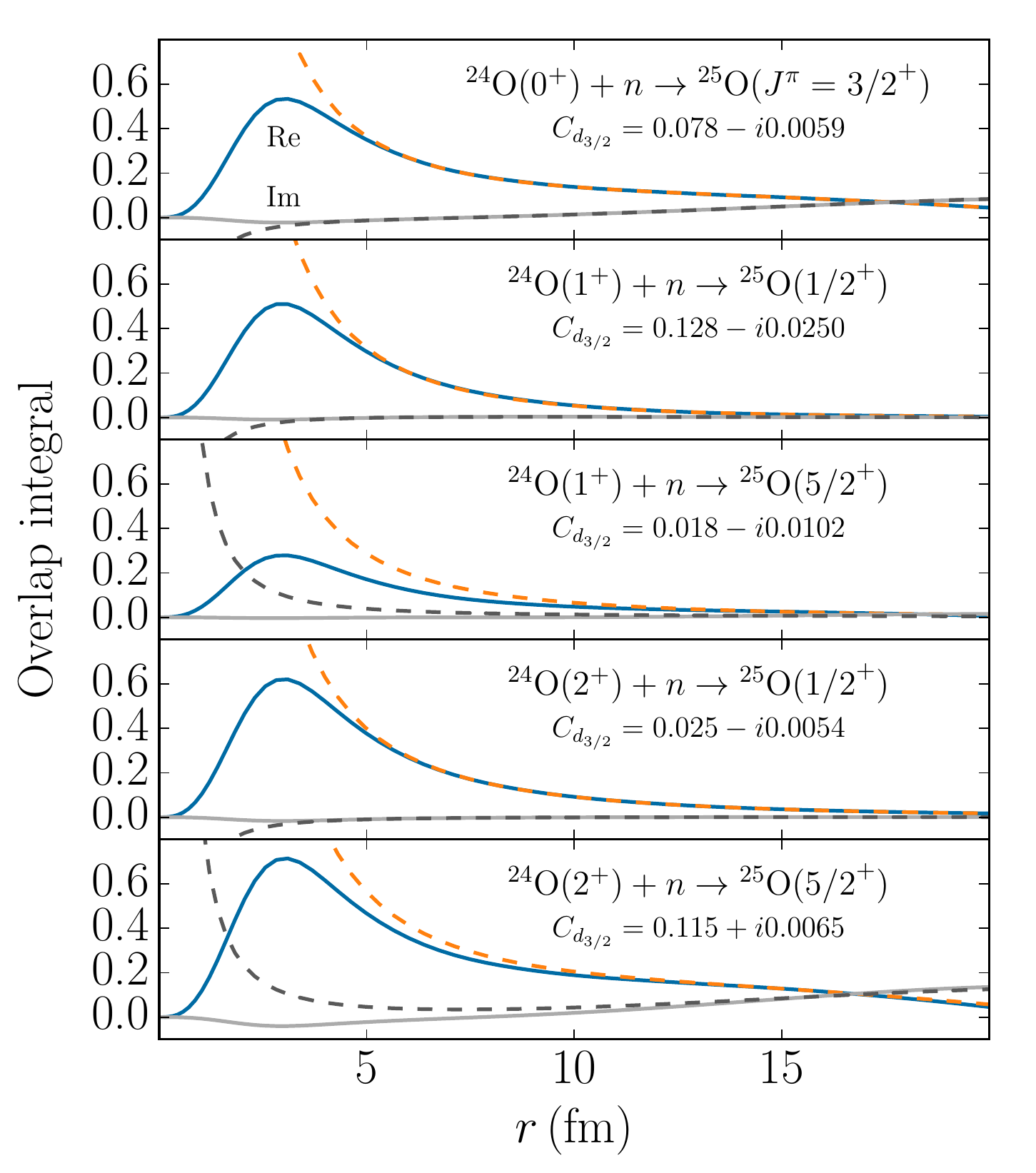}
    \caption{Overlap integrals for states in $^{24}$O and $^{25}$O (solid lines) from which the ANC values ${ {C}_{\ell j} }$  (in fm$^{-1/2}$) are extracted. The Whittaker functions used to fit the overlap integrals are marked by dashed lines. The ANCs for neutron transfer on $^{24}$O to negative parity states in $^{25}$O were too small ($\approx {10}^{-8}$ fm$^{-1/2}$) to show.}
    \label{fig5}
  \end{figure}
}

\section{Conclusions}
{
  \label{sec_conc}

  The anticipated excited states in $^{25}$O were investigated from both an experimental and theoretical point of view, 
  based on two assumptions: (\textit{i}) that excited states in $^{25}$O might have been missed by previous experiments using proton knockout on $^{26}$F 
  because of the expected structure of those states, 
  and (\textit{ii}) that strong couplings to the $fp$ continuum might strongly impact the structure of the negative parity excited states in $^{25,27}$O.

  On the experimental side, by using data from the neutron transfer reaction ${ {}^{24}\text{O} (d,p) {}^{25}\text{O} }$, 
  we extracted a limit on the relative cross section of a possible ${ {J}^{\pi} = {1/2}^{+} }$ state in $^{25}$O 
  and obtained $\sigma_{1/2+}/\sigma_{g.s.} = 0.25_{-0.25}^{+1.0}$ with respect to the ground state of $^{24}$O. 

  On the theoretical side, we studied the positive- and negative-parity excited spectrum of $^{25}$O within the GSM/DMRG framework by employing  an  effective two-body interaction optimized to experimentally known states of $^{23-26}$O. 
  By including couplings to the large $sdfp$ continuum space, we predicted excited states in $^{25}$O and $^{27}$O 
  and showed that the negative parity states go down  in energy when the number of neutrons increases.

  In summary, by critically analyzing the experimental and theoretical situation in neutron-rich oxygen isotopes, 
  we provided new insights about the possible presence of excited states in $^{25}$O, 
  and motivated further experimental studies employing neutron transfer reactions to study those states.
  Finally, we showed how the possible observation of negative parity states in $^{25}$O would hint into such states in $^{27}$O. 
}


\begin{acknowledgments}

  We would like to thank L.A. Riley for the use of the Ursinus College Liquid Hydrogen Target.
  This material is based upon work supported by the U.S.\ Department of Energy, Office of Science, Office of Nuclear Physics under Contract No. DE-AC02-05CH11231 (LBNL) and award numbers DE-SC0013365 (Michigan State University) and DE-SC0008511 (NUCLEI SciDAC-3 collaboration). This work was also supported by Department of Energy National Nuclear Securty Administration through the Nuclear Science and Security Consortium under award number DE-NA0003180 and by the National Science Foundation under award numbers PHY14-03906, PHY14-04236, PHY12-05537, PHY11-02511, PHY11-01745, PHY14-30152, PHY09-69058, PHY13-06074, and PHY12-05357.

\end{acknowledgments}

\bibliographystyle{apsrev4-1}
\bibliography{apsrev_refer}

\end{document}